\documentclass[twocolumn,showpacs]{revtex4}

\usepackage{dcolumn}
\usepackage{amsmath}
\usepackage{graphicx}
\begin{document}

\title{``Barchan'' Dunes in the Lab}

\author{Olivier Dauchot~$^1$,\ \ Fr\'ed\'eric Lech\'enault~$^1$,\ \ C\'ecile Gasquet~$^1$, Fran\c cois Daviaud~$^1$}
\affiliation{$~^1$ DSM/SPEC, CEA Saclay, 91 191 Gif sur Yvette, France}

\date{11 February 2002}

\begin{abstract}
We demonstrate the feasibility of studying dunes in a
laboratory experiment. It is shown that an initial sand pile, under a
wind flow carrying sand, flattens and gets  a shape recalling barchan
dunes. An evolution law is proposed 
for the profile and the summit of the dune. The dune dynamics is shown
to be shape invariant. The invariant shape, the ``dune function'' is
isolated.
\end{abstract}


\maketitle

Dunes dynamics has strong impact on the ecology and the economy of
sandy areas,  but remains far from being understood. 
Since the single major  work on sand dunes formation, written
by  R. A.  Bagnold in  1941~\cite{Bagnold},  a world  wide inventory  of
deserts has been developed in the fields by Sharp (1966)~\cite{Sharp}
and McKee (1979)~\cite{McKee}, in application of aerial photography by
Smith (1968)~\cite{Smith} and via Landsat imagery by Fryberger {\it et
al.} (1979)~\cite{Fryberger} and  Breed {\it et al.} (1979)~\cite{Breed},
among others. Five basic types of dunes have been recorded : 
crescentic, linear, star, dome and  parabolic. The most common dune is
the crescentic, also  called barchan. This type of  dune forms under
mono-directional winds. In that sense, barchans are also
the most "simple" dunes. They  are characterized by a crescentic crest
normal to  the wind  direction, with downwind  arms. The stronger the
wind, the  less  open   is  the   crescent.  The  windward
(resp. leeward) face is concave  (resp. convex). The barchans  move
over desert surfaces, while maintaining nearly constant shape.
Various models~\cite{Wipper,Werner,Nishimori,Galam,Hermmann}
have been proposed, from cellular automata to two-phase (static and moving sand) models.
However, if they  qualitatively capture most of the fields observations, they call
for more experimental data. On the one hand, field measurements 
are difficult to perform and often incomplete. On the other hand, 
it is believed that dunes have a minimal size of the order of the meter, 
not reducible to smaller laboratory scales.  

In this Note, we show experimentally that an initial sand pile, under a  wind
flow carrying sand, flattens and gets a shape recalling barchan dune. 
After  a  short  description  of our  experimental setup  and
protocol, we  present  first  quantitative results about the observed dune
patterns. We finally discuss why it is actually possible to
observe dunes in the lab.

Fields  observations  indicate  a   very  generic  behavior  of  dunes
essentially  controlled by the  wind direction  and force,  with little
dependence on  the details of  the wind structure. Accordingly  a very
simple   design   has  been   chosen   for   the   wind  tunnel.
It consists namely of two sections (figure 1(a)): the first one
($100\,{\rm mm}$ wide, $100\,{\rm mm}$ high,
$730\,{\rm mm}$ long) is devoted to establish a regular wind; the second
one ($230\,{\rm mm}$ wide, $175\,{\rm mm}$ high, $650\,{\rm mm}$ long) is open
at its end and its  floor is covered with $500\,{\rm \mu m}$ roughness
sand-paper. The wind speed is constant and set  below the "fluid
threshold" and above the "impact threshold". These thresholds are
defined as follows. For wind speeds above the "fluid threshold",
grains are picked up  from the surface and given a forward momentum,
before being brought back to the surface under their own weight, after
a typical "saltation length". If the surface is  covered with sand,
the grains loose most of their energy at the impact, but eject more
than one grain on average. Once the saltation is initiated, it is self-sustained as long as
the wind speed remains superior to the  "impact threshold". For the considered sand,
made up of mono-disperse $250\,{\rm \mu m}$ diameter  glass beads, the fluid
threshold $u_s^*\simeq 25\, {\rm cm/s}$ and the impact threshold
$u_c^*\simeq 20\, {\rm cm/s}$~\cite{Bagnold},  where $u^*$ is the friction velocity as
defined for a turbulent boundary layer~\cite{Tritton}.

\begin{figure}[htb]
\centering\includegraphics[width=8cm]{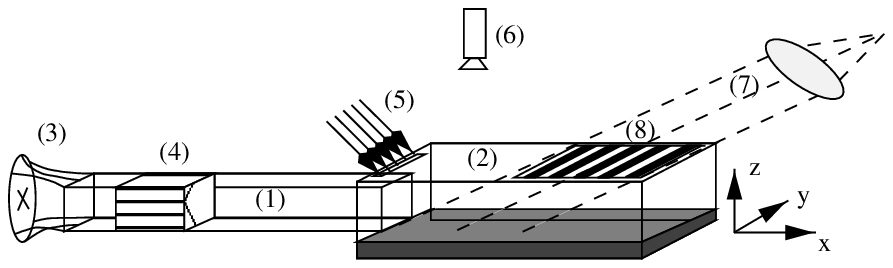}
(a)
\centering\includegraphics[width=8cm, height=6.5cm]{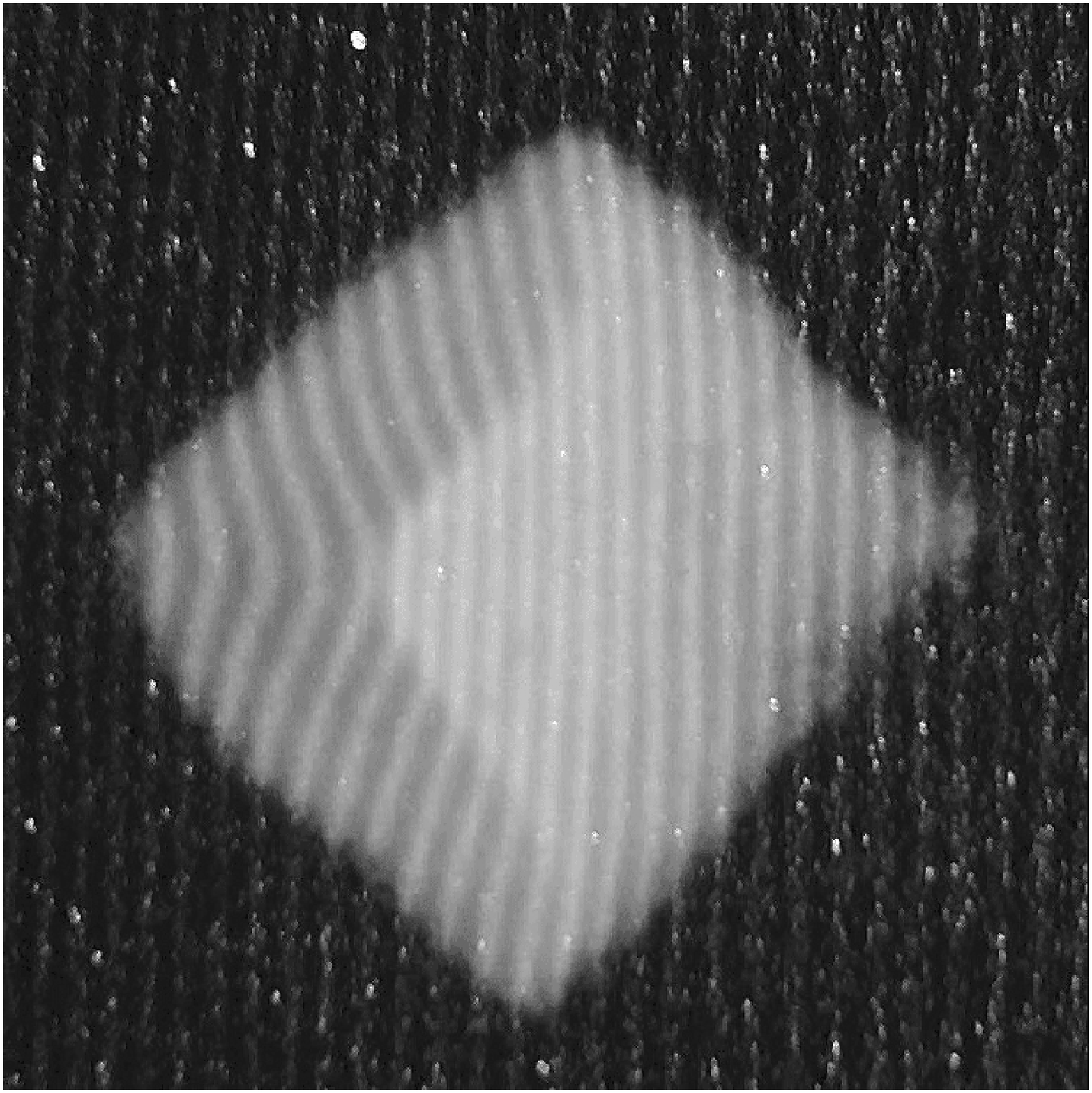}
(b)
\caption{set up and measurement technique (a)
Schematic drawing of the wind tunnel, with sand injection and
measurement set up : (1),(2) first and second wind tunnel sections;
(3) ventilator; (4) honeycomb; (5) sand injection; (6) CCD camera; (7)
parallel lighting; (8) light intensity modulating grid; (b) A typical
top view of a dune with the modulated lighting. The local streamwise
phase gradient is directly proportional to the local slope.}
\end{figure}

The initial  condition is a sand pile of volume $V_s$,
centered in the wind tunnel. The wind is then set up  and it is checked
that no sand motion occurs until a sand flux $q_s$ is added to the wind
at the top of the entrance of the second channel section; (in the
following $q_s$ is the  vertically integrated sand flow rate per spanwise
length unit). In the present study, we  report on three experiments
$E_{1,2,3}$ with respectively $q_{s_1}=1.25\pm 0.25\, {\rm g s^{-1} m^{-1}}$ and
$V_{s_1}=30\, {\rm cm}^3$, $q_{s_2}=q_{s_1}$ and $V_{s_2}=20\, {\rm cm}^3$, $q_{s_3}=5 \pm
0.25\, {\rm g s^{-1} m^{-1}}$ and $V_{s_3}=V_{s_1}$. The quantitative sand  pile
evolution, is obtained by profilometry : A sinusoidal light intensity
is projected onto  the experimental field and a CCD camera records the
field images from the top at  regular time interval. The local
streamwise phase gradient of the light intensity is  directly
proportional to the local slope (figure 1(b)). The recorded
images are processed in  order to obtain the topography $h(x,y)$.

\begin{figure}[htb]
\centering\includegraphics[width=8cm]{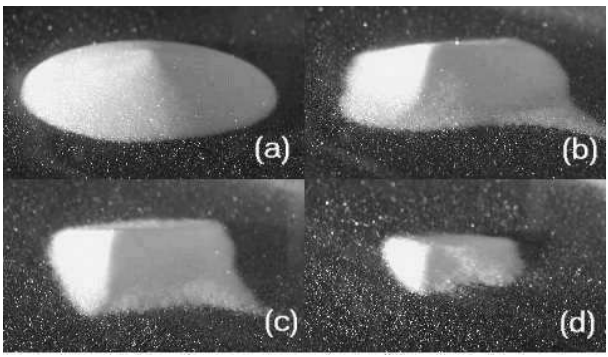}
\caption{A dune in the lab.  Typical evolution of the
initial sand pile (a) exposed to the wind charged in
sand coming from left upper corner of the picture. 
(a),(b),(c),(d) are separated by 5 minutes time intervals.}
\end{figure}

After a transient of the order of 5 minutes, the typical crescent
shape of the  dune (figure 2), with arms downwind and a 
slip-face on the leeward side appears.  A slight disturbance of this
face actually induces avalanching. The crescent crest is  rather open
as expected in low wind conditions. Under the present experimental
conditions, the dune is eroded until full removal of the initial
amount of sand. Figure 3  displays the evolution of the streamwise
profile $h^+(x,t)$ going through the dune summit.  The sand pile first
rapidly decreases in size to evolve towards a characteristic shape
with  a concave upwind side and a convex downwind side. Figure 3(b)
displays more frequent  time steps after the transient regime, scaled
by the summit abscissa $x_s$ and height $h_s$.  Once the dune profile is
reached, its shape remains the same up to rescaling. Figure 4 
displays the evolution of $x_s$ and $h_s$. The time has been scaled by
$t^*  =  \frac{\rho_s}{q_s}h_s(0)^2$,  where $\rho_s=1.57\,{\rm g/cm}^3$ is
the  sand bulk  density, $q_s$  the injected sand  flux  rate,  and
$h_s(0)$  the maximum height  at  the  initial condition. 

\begin{figure}[htb]
\centering\includegraphics[width=8cm]{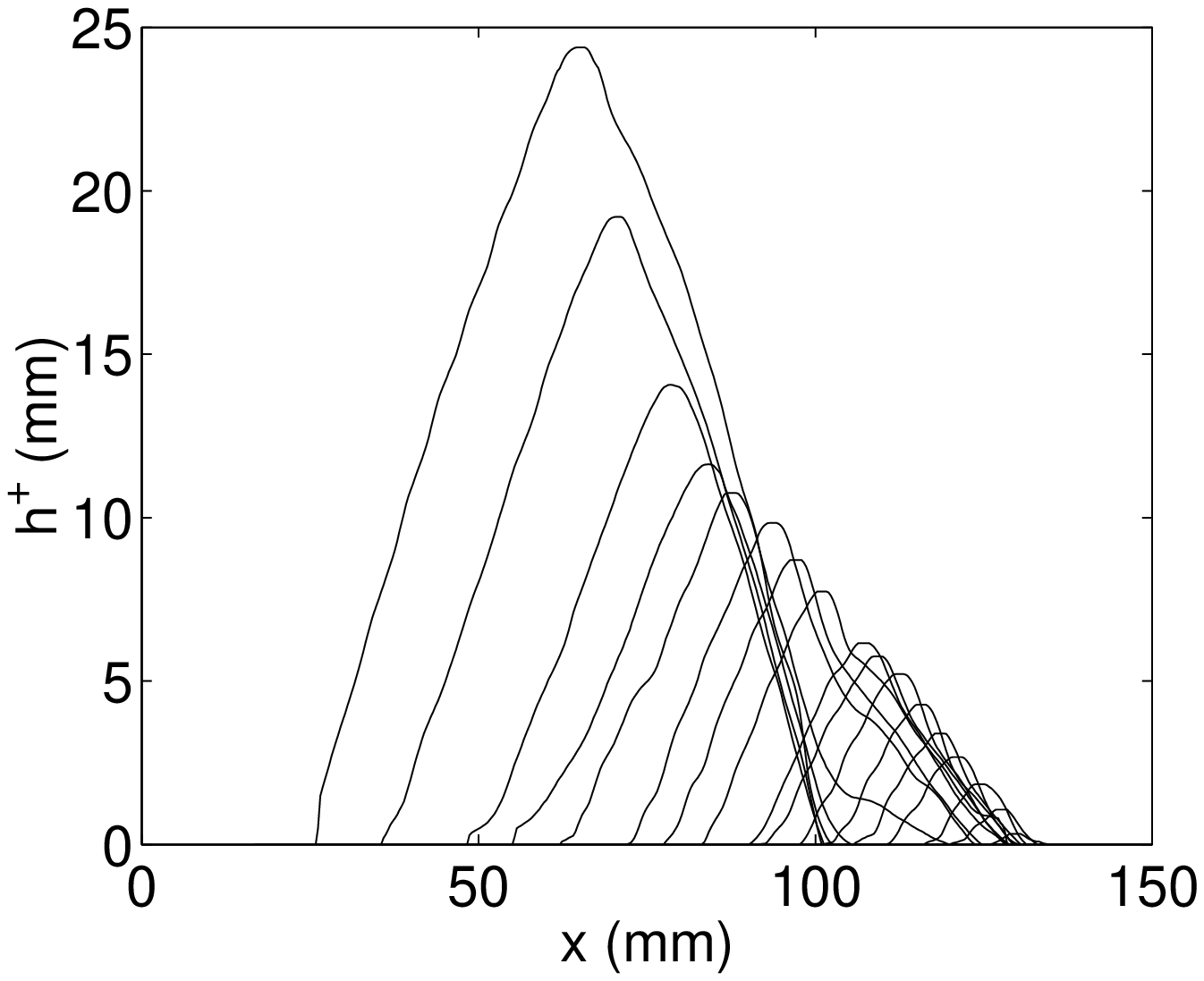}
(a)
\centering\includegraphics[width=8cm]{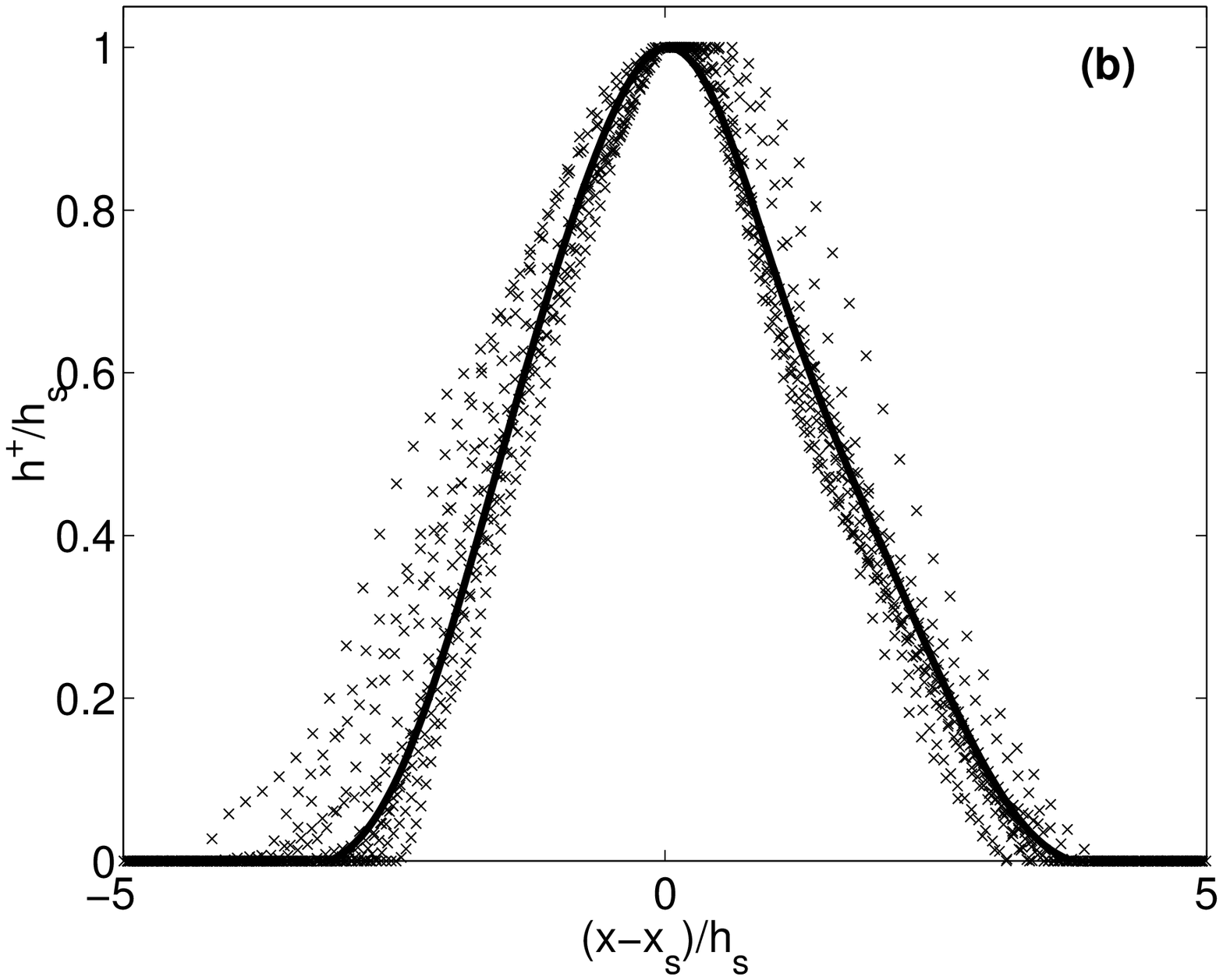}
(b)
\caption{Evolution of dune profile in experiment $E_2$. (a) From
initial condition to complete removal (time interval between curves is
2 min.) (b) Scaled profiles after transient (time interval between
dotted curves is 1 min.) The invariant shape , the  "dune function",
$\mathcal{D}$ is in dark line.} 
\end{figure}

This time scale $t^*$, here suggested by a dimensional analysis, 
is naturally recovered in the resolution of the mass balance equation (see later). 
The dunes obtained in the three  experiments exhibit the
same evolution, with identical scaled lifetimes. After the initial transient, 
the dunes propagate at constant velocity and their height decay like 
a square root of time, as shown in inset of figure 4(b). Altogether,  the dune
profile follows an evolution given by: 

$$
\begin{array}{ll}
& h^+(x,t) = h_s(t) {\mathcal{D}} (\frac{x-x_s}{h_s}),\\ 
& \\
{\rm with} & \frac{h_s(t)}{h_s(0)}=1-\alpha\sqrt{\frac{t}{t^*}}, \\ 
& \\
{\rm and} & \frac{x_s(t)-x_s(0)}{x_s(t^*)-x_s(0)} = \frac{t}{t^*}; \\
\end{array}
$$

\noindent 
where $\alpha $ is a dimensionless constant $\mathcal{D}$, and the
"dune function", is  the invariant shape of the dune profile. 

\begin{figure}[htb]
\centering\includegraphics[width=8cm]{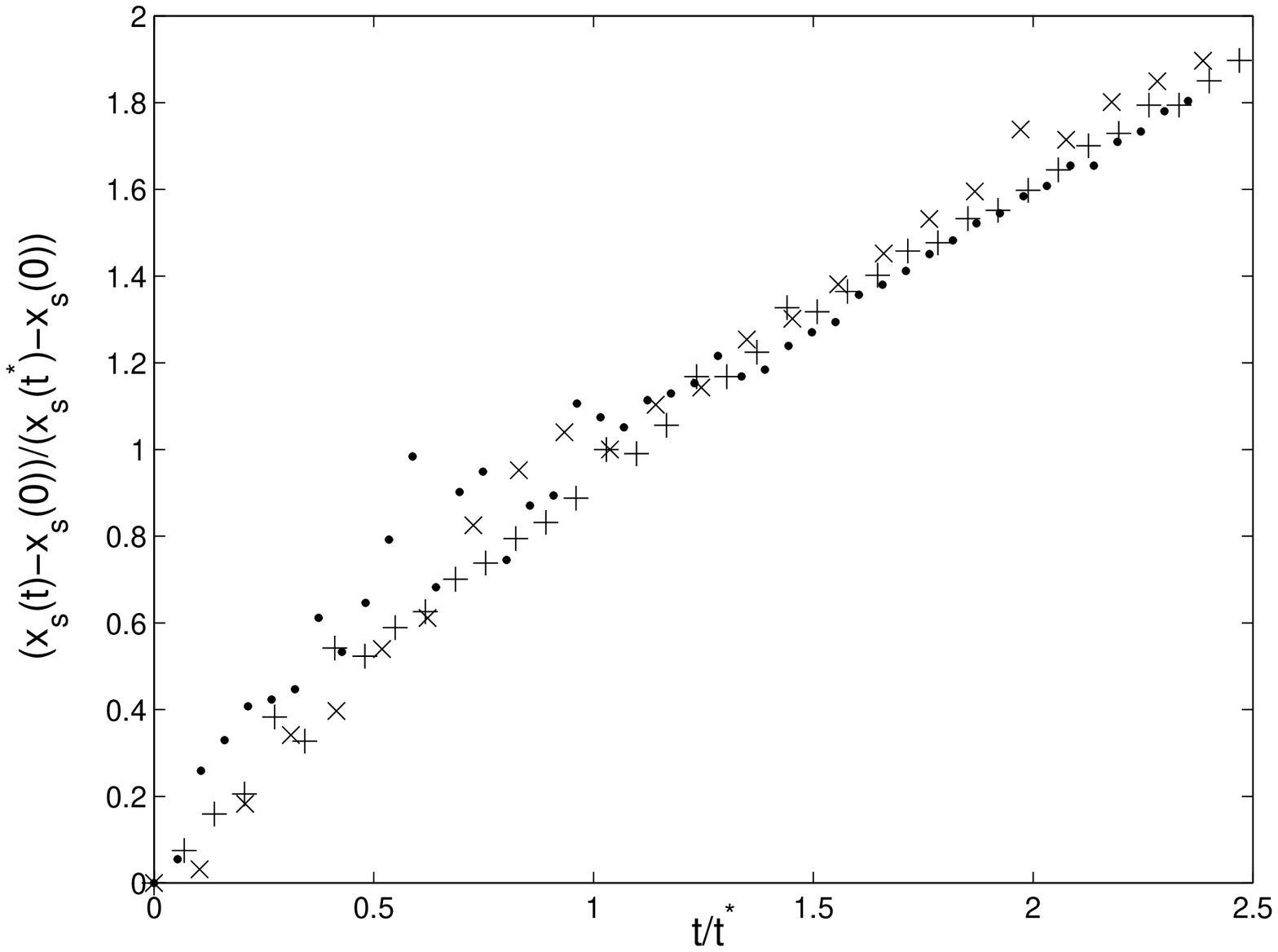}
(a)
\centering\includegraphics[width=8cm]{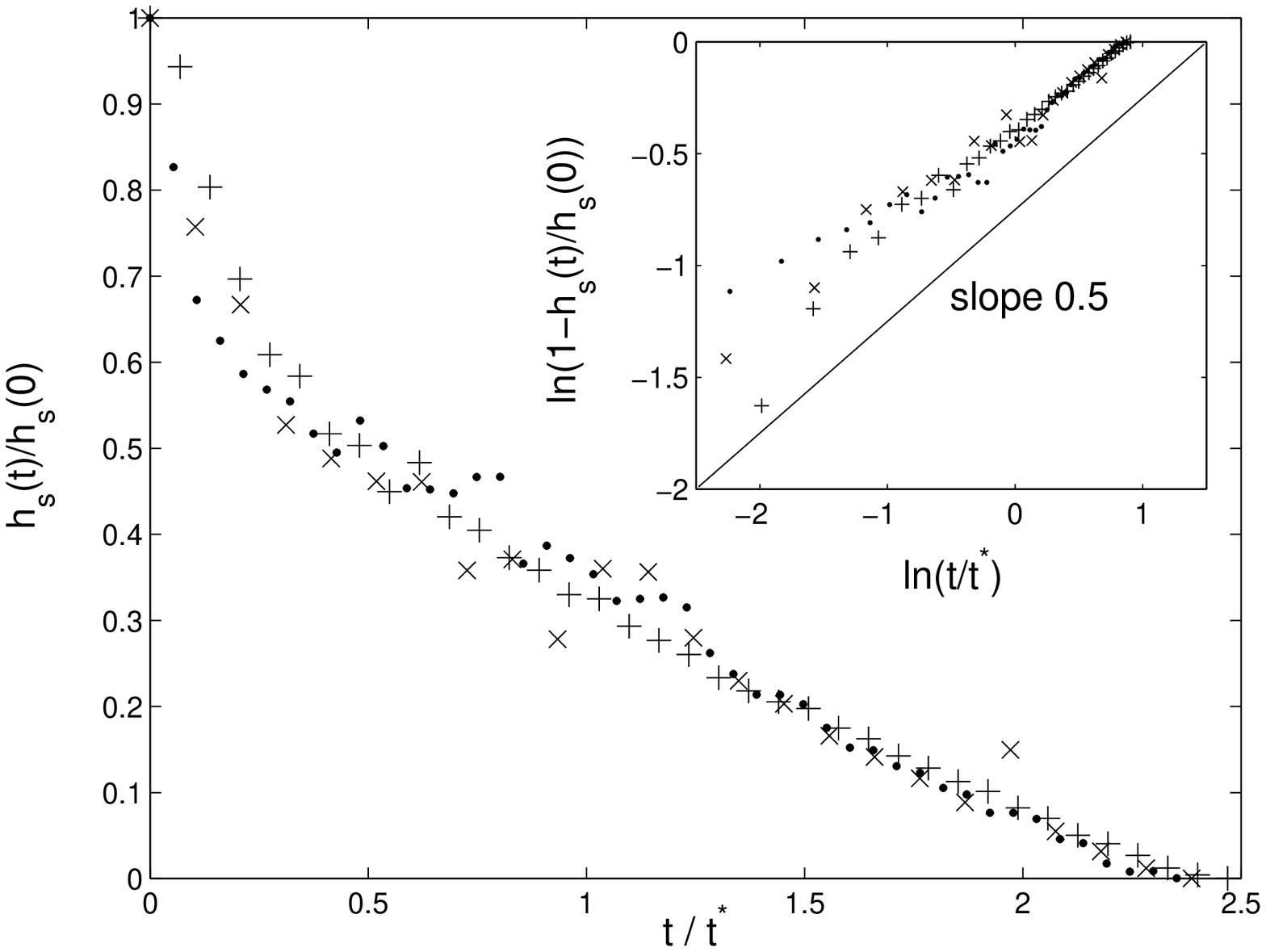}
(b)
\caption{Evolution of dune summit parameters. (a) summit
abscissa; (b) summit height with log-log plot in inset. Labels are $(\bullet)$
for experiment $E_1$, $(+)$ for $E_2$, $(\times)$ for $E_3$.} 
\end{figure}

These  evolution laws can easily be recovered by
considering mass transport and losses of the sand
pile. Let us consider an infinitesimal streamwise layer of the dune
with  profile $h^+ (x,t)$. We assume the airflow carrying  sand to be confined
to a layer of constant height $H$. We also assume the
sand flux to be uniform in height, so that the sand flux per height
unit is $\frac{q_s}{H}$. The expected proportionality between the
sand transport $\rho_s h_s v_s$ due to the dune motion at velocity
$v_s$ through a cross section at the abscissa of the maximum height,  
and the sand flux through a section of height $h_s$ outside the dune gives
$\rho_s h_s v_s \propto \frac{q_s}{H} h_s,$ thus the observed constant
velocity. Considering now the mass per spanwise length unit $M=A
\rho_s h_s^2$, where $A$ is the dimensionless area under $\mathcal{D}$, we
compute its decay rate. On one hand the balance of erosion and deposition rates
must be proportional to the sand flux per height unit times the dune
cross section. On the other hand, if the wind charged in sand is
confined in a layer of height $H$, when passing over the dune it is
accelerated by a factor of $H/(H-h_s)$ which may enforce the erosion, so that:

$$
\begin{array}{ll}
& \frac{dM}{dt}=2A \rho_s \frac{dh_{s}}{dt} h_s \propto
-\frac{q_s}{H}\frac{H}{H-h_s}h_s, \\
& \\
{\rm thus} & \frac{d\tilde{h_{s}}}{dt}(\tilde{H}-\tilde{h_s}) \propto -
 \frac{q_s}{\rho_s h_{s}^{2}(0)},
\end{array}
$$

\noindent
where $\tilde{h_s}=h_s/h_s(0)$; $\tilde{H}=H/h_s(0)$. Integrating with the
initial condition  $\tilde{h_s}(0)=1$, one has the relation:

$$\tilde{h_s}^2 -2\tilde{H}\tilde{h_s} +2\tilde{H} -1 =
\alpha^2\frac{t}{t^*},$$ 

\noindent
with $\alpha^2$ a positive proportionality constant. The
decreasing solution with time is:

$$\frac{h_s(t)}{h_s(0)}=\tilde{H}-\sqrt{\alpha^2\frac{t}{t^*}+(\tilde{H}-1)^2}.$$

\noindent 
The simpler expression obtained experimentally is recovered when
$\tilde{H}=1$, in agreement  with the visual observation that the
injected sand flux extends on a height of the same  order of the
initial sand pile. This specific feature may be different in
nature.

How is it that we could observe dunes at a much smaller scale than in
nature {\sl with the same kind of sand}? In deserts, dunes have to grow from
flat initial conditions. As  already pointed out by
R. A. Bagnold~\cite{Bagnold}, the self accumulation process
responsible for  dunes building is efficient if the wind speed is high
enough to charge the wind in sand  prior to the dunes field. For a
typical wind speed of $25\, {\rm km/h}$, at one meter above ground,
the friction velocity is $u^* \simeq 1\, {\rm m/s}$ and the saltation 
length is $l_s \simeq 10\, {\rm cm}$~\cite{Bagnold}. A typical  dune streamwise
extension is of the order of a hundred to a thousand times the
saltation  length. Such ratios are clearly out of reach of the lab
experiment. Yet the above results  clearly demonstrate the
feasibility of dune investigation in a lab, at least during a long transient.
Moreover, the  downwind
motion of the dune reveals a transport mechanism similar to the one
observed in nature, in which the windward face is eroded, while the
leeward face  accumulates sand until avalanches set up. Having dealt with
low wind, artificially charged in sand, we have lowered the saltation
length down to $l_s \simeq 1\, {\rm cm}$. This is more than one  tenth
of the sand pile streamwise extension, but it is small enough to let
saltation occurs  on the windward face of the dune and accordingly to
let the basic dynamics mechanisms  take place. Since there must
actually be not just a single saltation length but a  distribution of
them around an average, the main difference between nature and
our experiment is that, in nature saltation jumps an order of
magnitude larger than the  averaged value still transport sand on the
dune, whereas in our experiment, the grains  are lost for the
dune. This difference may explain why our experimental dunes 
erode so fast. 

Altogether,  we believe that the  initial sand pile  evolves according
to the  same elementary erosion and  deposition mechanisms as  those
involved in  a stationary dune dynamics in nature. Together with the
robustness of the dune shape underlined by Werner's elementary model~\cite{Werner},
it leads us to conclude to the relevance of the dune function obtained
here. Whether the laboratory dunes are exactly barchan dunes, in a
sense which should be precised, needs further investigation of both field
and experimental data. Beyond  the above quantitative  results, the present work has
proved  the feasibility of investigating small dunes convenient for
lab investigations. It calls for
further  studies  under  various  wind conditions,  with  different  sand
types and opens a  new kind of investigations in desert studies.  

%%%%%%%%%%%%%%%%%%%%%%%%%%%%%%%%%%%%%%%%%%%%%%%%%%%%%%%%%%%%
%%%  Acknowledgements  %%%
%%%%%%%%%%%%%%%%%%%%%%%%%%
We thank D. Bonamy, E. Bertin, for helpful
discussions and P. Meininger for technical support.
%%%%%%%%%%%%%%%%%%%%%%%%%%%%%%%%%%%%%%%%%%%%%%%%%%%%%%%%%%%%
%%%  Bibliography  %%%
%%%%%%%%%%%%%%%%%%%%%%%

%

\end {document}